\documentclass[prb,aps,showpacs,reprint,twocolumn,floatfix]{revtex4-1}

\usepackage{amsmath}
\usepackage{graphicx}
\usepackage{amsfonts}
\usepackage{float}

\begin{document}

\title{Strong In-plane Anisotropy in the Electronic Properties of Doped Transition Metal Dichalcogenides exhibited in W$_{1-x}$Nb$_{x}$S$_{2}$}

\author{Siow Mean Loh}
\email{s.loh@warwick.ac.uk}
\thanks{These authors contributed equally.}
\affiliation{Department of Physics, University of Warwick, Coventry CV4 7AL, United Kingdom.}

\author{Xue Xia}
\email{x.xia@warwick.ac.uk}
\thanks{These authors contributed equally.}
\affiliation{Department of Physics, University of Warwick, Coventry CV4 7AL, United Kingdom.} 

\author{Neil R. Wilson}
\email{neil.wilson@warwick.ac.uk}
\affiliation{Department of Physics, University of Warwick, Coventry CV4 7AL, United Kingdom.} 

\author{Nicholas D. M. Hine}
\email{n.d.m.hine@warwick.ac.uk}
\affiliation{Department of Physics, University of Warwick, Coventry CV4 7AL, United Kingdom.} 

\date{\today}

\begin{abstract}
We study the electronic properties of monolayer transition metal dichalcogenide materials subjected to aliovalent doping, using Nb-doped WS$_{2}$ as an exemplar. Scanning transmission electron microscopy imaging of the as-grown samples reveals an anisotropic Nb dopant distribution, prompting an investigation of anisotropy in electronic properties. Through electronic structure calculations on supercells representative of observed structures, we confirm that local Nb-atom distributions are consistent with energetic considerations, although kinetic processes occurring during sample growth must be invoked to explain the overall symmetry-breaking. We perform effective bandstructure and conductivity calculations on realistic models of the material that demonstrate that a high level of anisotropy can be expected in electronic properties including conductivity. In-plane anisotropy of the conductivity is predicted to be as high as 5:1, which is higher than previously observed in any TMDC system in the [Mo,W][S,Se]$_2$ class.
\end{abstract}


\maketitle

\section{Introduction}

The varied properties of transition metal dichalcogenides (TMDCs) have given these novel materials a central role in the rapidly-growing field of two-dimensional materials. They show promise in applications including electronic devices, nanoelectronics and optoelectronics \cite{Tunable_Band_Gap, Quaternary,Spin-orbit_engineering,Transistors_NbxW1-xS2_monolayers,MoS2_transistors,Ghazaryan2018,Ominato2020,Bhowal2020}. The specific category of TMDCs of the form MX$_{2}$ where M$=$Mo, W and X$=$S, Se has been particularly widely-studied, and the high quality electronic properties these materials exhibit are increasingly being integrated into devices\cite{MoS2_transistors,Single-Layer_WS2,WS2_Transistors,WSe2_p-FETs,WSe2_FETs,Single-Layer_WSe2,MoSe2_FETs,MoSe2_FETs_on_Parylene-C_Substrate,Liu2020}. For example, current on/off ratios of order 10$^{8}$ have been achieved in MoS$_{2}$ and WSe$_{2}$ monolayers. High mobilities, $\sim$200 cm$^{2}$ V$^{-1}$ s$^{-1}$ and 83 cm$^{2}$ V$^{-1}$ s$^{-1}$ can also be realized for MoS$_{2}$ monolayer and WSe$_{2}$ monolayer
respectively\cite{MoS2_transistors,WSe2_FETs}. Furthermore, the ambipolar behavior of MX$_{2}$ paves the way to application in field-effect transistors (FETs)\cite{Single-Layer_WSe2,MoSe2_FETs}. To achieve more functional TMDCs, the controllable-introduction of charge carriers is widely applied.

Two-dimensional material make feasible several novel ways to introduce carriers (eg. gating and adatoms), but the traditional approach, as in the semiconductor industry, remains substitutional doping with acceptor or donor impurities. Analogy to use of B as an acceptor and P as a donor for silicon\cite{B_doped_Si,B_doped_Si_rapid_thermal_proecessing,Heavily_Doped_Silicon}, suggests (Nb, Ta) as hole donors\cite{Transistors_NbxW1-xS2_monolayers,Nb_doped_MoSe2,Gao2019} and (Tc, Re) as electron donors\cite{Rh_doped_MoSe2,Wang2019} for MX$_{2}$ devices. Aliovalent dopant impurities can be expected to have strong effective interactions between impurity atoms that may influence dopant distributions.

Doping levels as high as 15\%  mean individual dopant atoms cannot be considered isolated, which raises the questions of how they are distributed, how this depends on the growth conditions and synthesis approaches, and how this distribution influences other properties\cite{Synthesis_2D_TMDCs}. Our previous studies of alloy TMDCs with atomic-resolution imaging have observed an effectively random distribution of dopants for highly-compatible TM atoms, as in Mo$_{1-x}$W$_{x}$S$_{2}$ alloys\cite{Xia2021}. Conversely, Azizi et al. produced nonrandom and vibrationally anisotropic stripe patterns \cite{Atomically_Thin_Stripes_TMDC_monolayer} which ran parallel to the edges of the triangular flakes in monolayer Mo$_{1-x}$W$_{x}$S$_{2}$ around 50\% doping. Their appearance was considered to be a kinetic effect and is also affected by the fluctuations related to the local chemical potentials of transition metal atoms and chalcogen atoms. Azizi et al. later observed atomic ordering in Nb$_{x}$Re$_{1-x}$S$_{2}$ resulting from competition between geometrical constraints and nearest-neighbor interactions \cite{Frustration_and_Ordering_in_NbReS2}.

Such patterns hold promise in the creation of controllable levels of anisotropy. Properties previously observed to show anisotropy in low-dimensional materials include mechanical, electronic and thermoelectric transport \cite{In_plane_anisotropic_2D,Highly-anisotropic_SnSe,Penta_PdX2}. To date, this anisotropy has mostly resulted directly from reduced symmetry of the crystal structure, such as the distortions observed in transition metal tellurides. Controllable bottom-up synthesis of materials with high overall symmetry but built-in anisotropy in their local composition is a very promising alternative that could lead to considerably greater degrees of tunable anisotropy \cite{Synthesis_Transport_Nb-Doped_WS2, W1-xNbxS2_CVD-1}.

In this work we study W$_{1-x}$Nb$_{x}$S$_{2}$: considering this as an alloy of a metal (NbS$_{2}$) and a semiconductor (WS$_{2}$), we can expect significant deviations from the behaviour of either parent material. At very low $x$, a  metal-semiconductor transition can be expected as the doping concentration decreases, while at higher Nb concentration, clustering may occur, depending on the details of the Nb-Nb interactions. Previous studies of W$_{1-x}$Nb$_{x}$S$_{2}$ with atomic-resolution imaging have used samples with relatively low doping levels, eg in Gao et al. \cite{Doping_in_TMDCs} where $x\sim 6.7\%$ is seen. By contrast we focus on a higher doping level, $x\sim 10\%$, where Nb-Nb interactions become more prevalent.

We are motivated by surprising results from our growth of W$_{1-x}$Nb$_{x}$S$_{2}$ samples, which on examination in scanning transmission electron microscopy (STEM) show a high degree of anisotropy in the distribution of Nb atoms (Fig. 1(a)).
Nb atoms appear to cluster into aligned line segments: our goal in this work is therefore to predict and explain what consequences this has in terms of anisotropy in the electronic properties.

Clear evidence of a semi-regular striped pattern is seen, which we find corresponds to single-atom stripes predominantly comprising NbS$_{2}$, spaced by around 4-5 unit cells comprising predominantly WS$_{2}$. We perform simulations to understand three key properties that combine to influence the electronic behaviour of 2D alloy systems. First we investigate the energetic ordering of chosen motifs for dopant distribution. We study exemplar arrangements of Nb atoms, to test whether these structures are energetically feasible. We then calculate the electronic bandstructure of this material with a range of dopant distributions, to show that the electronic properties are sensitive to the anisotropy in the distribution of Nb dopants. Finally, we calculate conductivities of these model structures and estimate the level of in-plane anisotropy in the conductivity that can be achieved in realistic systems.

\begin{figure*}[bt]
  \centering
  \includegraphics[width=\textwidth]{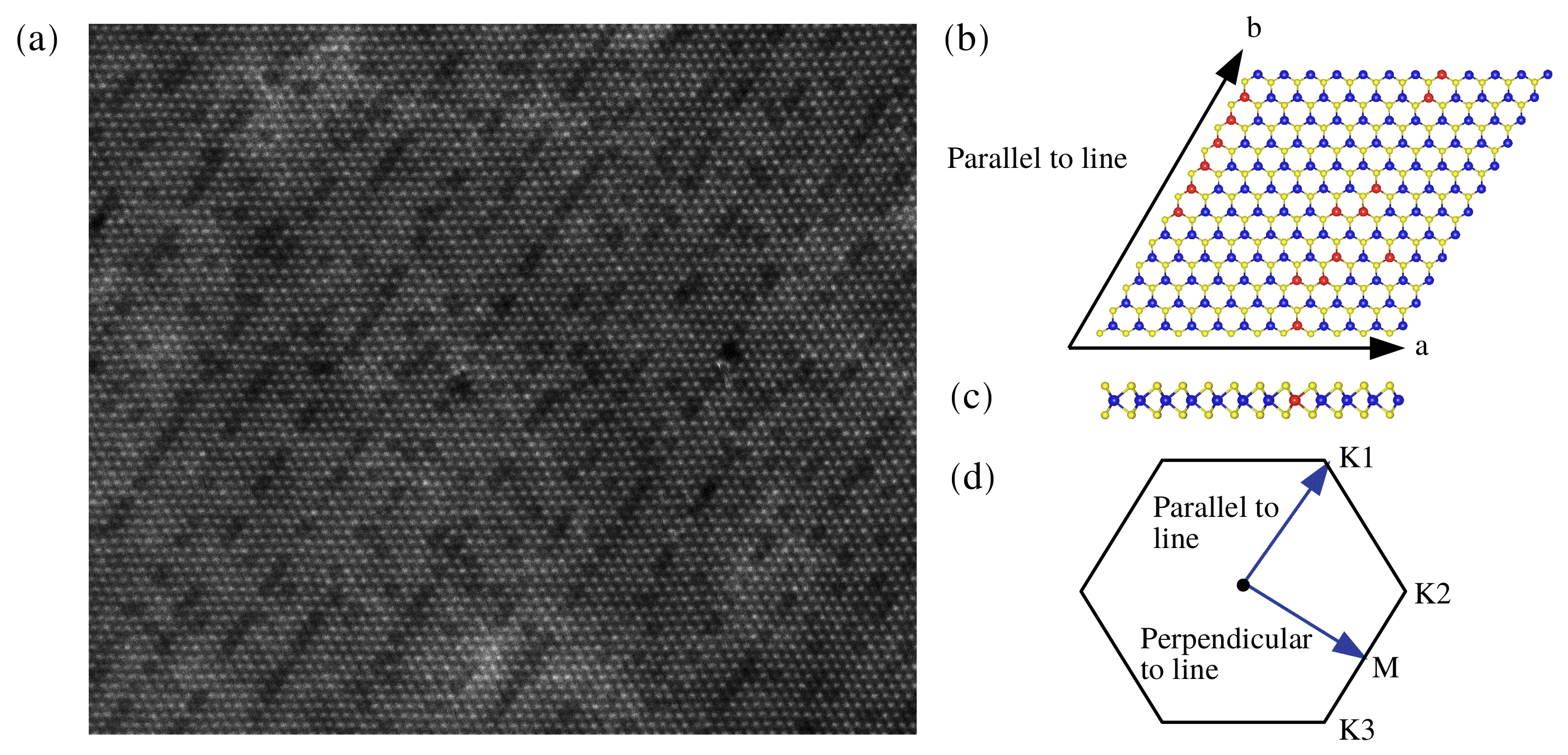}
\caption{ (a) STEM image for W$_{0.9}$Nb$_{0.1}$S$_{2}$ monolayer (bright: W, dark: Nb). (b) Top view and (c) side view of atomic structure from a region of (a), where the $\textbf{a}$-axis and $\textbf{b}$-axis are indicated. (Red: Nb, Blue: W, Yellow: S) (d) First Brillouin zone of the unit cell with some specified k-points indicated. The arrows show the directions parallel and perpendicular to the line.}
\end{figure*}

\section{Experimental methodology}

Material synthesis: W$_{0.9}$Nb$_{0.1}$S$_2$ single crystals were synthesised via chemical vapour transport (CVT) following similar protocols to previous reports \cite{Romanenko2017,Xia2021}. First, powders of the pure elements Nb (purity 99.99$\%$, Sigma Aldrich), W (purity 99.99$\%$, Sigma Aldrich) and S (purity 99.99$\%$, Sigma Aldrich) were mixed stoichiometrically and sealed in a quartz ampoule. The mixture was heated to 1000 $^{o}$C and maintained at that temperature for 3 days, forming a high-quality alloy compound. This compound was transferred to a new quartz ampoule and iodine (purity 99.99$\%$, Sigma Aldrich) was added. The ampoule was evacuated to $10^{-6}$ mbar in ice and sealed. The quartz ampoule was placed in a three-zone furnace for crystal growth by CVT. The charge zone (with the alloy powder) was slowly heated to 1050 $^{o}$C and kept at this temperature for 20 days, with the growth end at 950 $^{o}$C. After the ampoule was cooled down to room temperature, the ampoule was opened in air and the as-grown crystals were collected from the growth end of the ampoule.

Atomic resolution imaging: to enable atomic resolution imaging of the monolayer structure by STEM, monolayer W$_{0.9}$Nb$_{0.1}$S$_2$ flakes were encapsulated in graphene to increase their stability under the electron beam, as reported previously for Mo$_{1-x}$W$_x$S$_2$ samples \cite{Xia2021}. The W$_{0.9}$Nb$_{0.1}$S$_2$ single crystals were mechanically exfoliated onto chemical vapour deposition (CVD) grown graphene on copper and subsequently a layer of polycarbonate (PC) coated CVD-grown graphene was transferred on top. The copper underneath was etched away using ammonia persulphate, then the PC supported stack was gently washed via several fresh batches of deionised water. The resultant graphene encapsulated W$_{0.9}$Nb$_{0.1}$S$_2$ flakes were wet transferred onto a fresh TEM grid and the PC film washed away with acetone.
STEM analysis was performed in a JEOL ARM200F TEM with CEOS probe and image aberration correction, at 80 kV acceleration voltage. The annular dark field (ADF) images were recorded at a probe current of ~23 pA and a convergence semi-angle of ~25 mrad via a JEOL annular field detector with an inner and outer collection semi-angle of 45 and 180 mrad, respectively \cite{Huang2014,Sanchez2016}. The scanning rate was typically 10 $\mu$s per pixel, each image consists of 4096 $\times$ 4096 pixels. 

\section{Theoretical Background and Computational Methods}

The total energy obtained from a density functional theory calculation on a model system can be used as a substitute for the free energy (disregarding vibrational effects) that would determine the likelihood of encountering such a structure at thermal equilibrium. However, multiple factors indicate that in fabricating this particular alloy crystal by CVT method we obtain structures determined by the kinetics of the growth process\cite{Atomically_Thin_Stripes_TMDC_monolayer}, rather than distributions representing thermal equilibrium. One key indication is that there is a symmetry-breaking, presumably provided by the growth front during the CVT process, which leads to the formation of line segments of dopants, all aligned with each other over the lengthscale of an entire flake, of micrometer size.

Within such a non-equilibrium kinetic process, multiple factors such as temperature, relative ratio of the precursors, growth rate and some intrinsic parameters that influence the activation energy are all capable of producing different growth morphologies\cite{Kinetic_Monte_Carlo_simulation}. We therefore do not seek to determine from our calculations what dopant distribution motifs would be present for an equilibrated structure, but rather take inspiration from the observed distributions, as illustrated in Fig. 1(a), and study idealised models of chosen motifs. This is not intended to be an exhaustive coverage of the motifs, but rather an illustrative selection of those that are present.

To determine the feasibility of the structures observed from the point of view of energetics, we define the formation  energy for a specific W$_{1-x}$Nb$_{x}$S$_{2}$ alloy supercell labelled $S$, relative to the perfect WS$_{2}$ monolayer, as\cite{Defect_formation_energies_in_GaAs}:
\begin{equation}
\Delta E_{\mathrm{F}}(S) = E_{\mathrm{T}}(S) - E_{\mathrm{T}}^{\:\mathrm{perf}} + \sum_{i} m^{S}_i \mu_{i}
\end{equation}
where $\Delta E_{\mathrm{F}}(S)$, $E_{\mathrm{T}}(S)$ and $E_{\mathrm{T}}^{\:\mathrm{perf}}$ are the formation energy, total energy of the alloy model supercell $S$, and the total energy of the corresponding WS$_{2}$ monolayer supercell, respectively. $\mu_i$ is the chemical potential of species $i$ (Nb or W) and $m_{i}(S)$ is the number of substituted atoms of species $i$ in model $S$.

Taking this model supercell to represent a cluster of dopant atoms, we can then define a binding energy of this cell relative to the corresponding number of isolated dopant atoms:
\begin{align}
E_{\mathrm{B}} &= E_{\mathrm{F}}(m \mathrm{Nb}) - m E_{\mathrm{F}}(1 \mathrm{Nb})
\nonumber\\
&= E_{\mathrm{T}}(m \mathrm{Nb}) - m E_{\mathrm{T}}(1 \mathrm{Nb}) + (m - 1)E_{\mathrm{T}}^{\:\mathrm{perf}}
\end{align}
where $E_{\mathrm{B}}$, $E_{\mathrm{T}}$($\textit{m}$Nb) and $E_{\mathrm{T}}$(1Nb) refer to the binding energy, total energy of $m$ substituted Nb atoms and 1 substituted Nb atom, respectively. 

We use the functionality available in ONETEP for calculating effective bandstructure \cite{Spectral_function,Spectral_function_in_ONETEP} to obtain the spectral function for models of the W$_{1-x}$Nb$_{x}$S$_{2}$ monolayer. Previously applied to TMDC and post-transition metal chalcogenide heterostructures \cite{Spectral_function_in_ONETEP, two-dimensional_semiconductor_heterostructures, ghost_anti-crossing_two-dimensional_heterostructures}, this spectral function unfolding method is also appropriate to the bandstructures of disordered alloys. The spectral function within the Brillouin zone of the underlying primitive cell is defined as \cite{Spectral_function, Tunable_Band_Gap}
\begin{equation}
A(\vec{k_{i}},E) = \sum_{m, n} \vert \langle \vec{K}m \vert \vec{k_{i}}n \rangle \vert^{2} \delta(E_{m}-E)
\end{equation} 
where $\vert \vec{K}m \rangle$ is a supercell eigenstate with wavevector $\vec{K}$ and band index $m$, and $\vert \vec{k}_{i}n \rangle$ is a `primitive cell' eigenstate formed by unfolding from the BZ of the supercell. This represents projection of a set of eigenstates $\vert \vec{k}_{i}n \rangle$ from primitive cell to the eigenstates from supercell $\vert \vec{K}m \rangle$ at energy $E$.

We utilize a semi-classical Boltzmann Theory\cite{Boltztrap} to obtain the conductivity of the alloy models. The conductivity tensor is given by
\begin{align}
\sigma_{\alpha\beta}(\varepsilon) &= \frac{1}{N} \sum_{i,\vec{k}} e^{2} \tau_{i,\vec{k}} \: v_{\alpha}(i,\vec{k}) \: v_{\beta}(i,\vec{k}) \: 
\nonumber\\
& \frac{\delta(\varepsilon-\varepsilon_{i,\vec{k}})}{d\varepsilon}
\end{align} 
where $N$ and $e$ are the number of k-points and the electron charge, respectively. $v_{\alpha}$ and $v_{\beta}$ are the group velocities, given by
\begin{equation}
v_{\alpha}(i,\vec{k}) = \frac{1}{\hbar} \: \frac{\partial \varepsilon_{i,\vec{k}}}{\partial k_{\alpha}}
\end{equation} 
The transport tensor is calculated as
\begin{align}
\sigma_{\alpha\beta}(T;\mu) &= \frac{1}{\Omega} \int \sigma_{\alpha\beta}(\varepsilon) 
\nonumber\\
& \left[-\frac{\partial f_{\mu}(T;\varepsilon)}{\partial \varepsilon} \right] d\varepsilon
\end{align}
where $\Omega$, $\mu$ and \textit{f$_{\mu}$}(T; $\varepsilon$) refer to the cell volume, electron chemical potential and the Fermi-Dirac distribution, respectively.

We make use of three simulation packages for the electronic structure calculations: for smaller supercells, we use the plane-wave DFT code CASTEP\cite{CASTEP}. This was first employed to calculate the optimized lattice constant of WS$_{2}$ (3.19 \AA), NbS$_{2}$ (3.35 \AA) (consistent with other experimental and theoretical results\cite{Quantum_confinement_in_TS2,TMDCs}) and of the alloys. Calculations to determine the energetic preference of different dopant arrangements are also performed in CASTEP to ensure results that are very well-converged with respect to basis size. In CASTEP, a fixed cut-off energy of 500 eV is used throughout, and the Monkhorst-Pack k-point grids are chosen such that no spacing was less than 0.11 \AA$^{^{-1}}$. On the fly generated ultrasoft pseudopotentials \cite{Vanderbilt_1990} are used. The supercell sizes used are chosen to balance the competing demands of avoiding finite size errors due to spurious periodicity, and keeping the computational cost within feasible limits. The total energy was converged within 1 $\times$ 10$^{-5}$ eV/atom and 2 $\times$ 10$^{-5}$ eV/atom for electronic minimization and geometry optimization tasks, respectively.

Spectral function unfolding bandstructure calculations are performed with the ONETEP linear-scaling density functional theory (LS-DFT) package\cite{ONETEP-1, ONETEP}. We use a cut-off energy of 1600 eV and a NGWF radius of 13.0 Bohr, and projector-augmented wave (PAW) \cite{PAW_in_ONETEP} datasets from the GBRV library \cite{GBRV_library}.

In both CASTEP and ONETEP calculations, the generalized gradient approximation (GGA) in the scheme of the Perdew Burke and Ernzerhof (PBE) functional\cite{GGA} is used as the exchange correlation potential. Given that for any $x>0$ this material is metallic, the ability of the functional to predict bandgaps is not relevant in this case. Previous studies have shown that in the $x=1$ parent compound NbS$_2$, both Coulomb and electron–phonon interactions can individually have a strong effect\cite{vanLoon2018}, but their combined effect produces a bandstructure that closely resembles the non-interacting band structure, so we do not consider these interactions explicitly in the current work. A vacuum spacing of 20 \AA \: is used to avoid spurious interaction between isolated layers of W$_{1-x}$Nb$_{x}$S$_{2}$. These parameters were proven to be sufficient to satisfy the convergence criteria.

For calculations of the conductivity, the BoltzTrap \cite{Boltztrap} code was utilized as a post-processing package, applied to the results output by the CASTEP package. To enable spin orbit coupling, we utilise, within CASTEP, $j$-dependent pseudopotentials \cite{j-dependent_pseudopotential} obtained from CCPForge. The integration in equation (6) is performed on a grid with d$\varepsilon$ = 0.0068 eV. The number of lattice points is 5 times the number of k-points. A 8$\times$8$\times$1 supercell was adopted in the calculation, with the spectral k-point mesh set to 6$\times$6$\times$1. Note that the unknown relaxation time $\tau$, in terms of which the results are given, is approximated as a constant common to all calculations.

All figures of the atomic structures and the isosurfaces of charge density (see supplementary material) shown in this work were displayed through vizualisation for electronic and structural analysis (VESTA) software \cite{VESTA}.

\begin{figure*}[hbt]
  \centering
  \includegraphics[width=\textwidth]{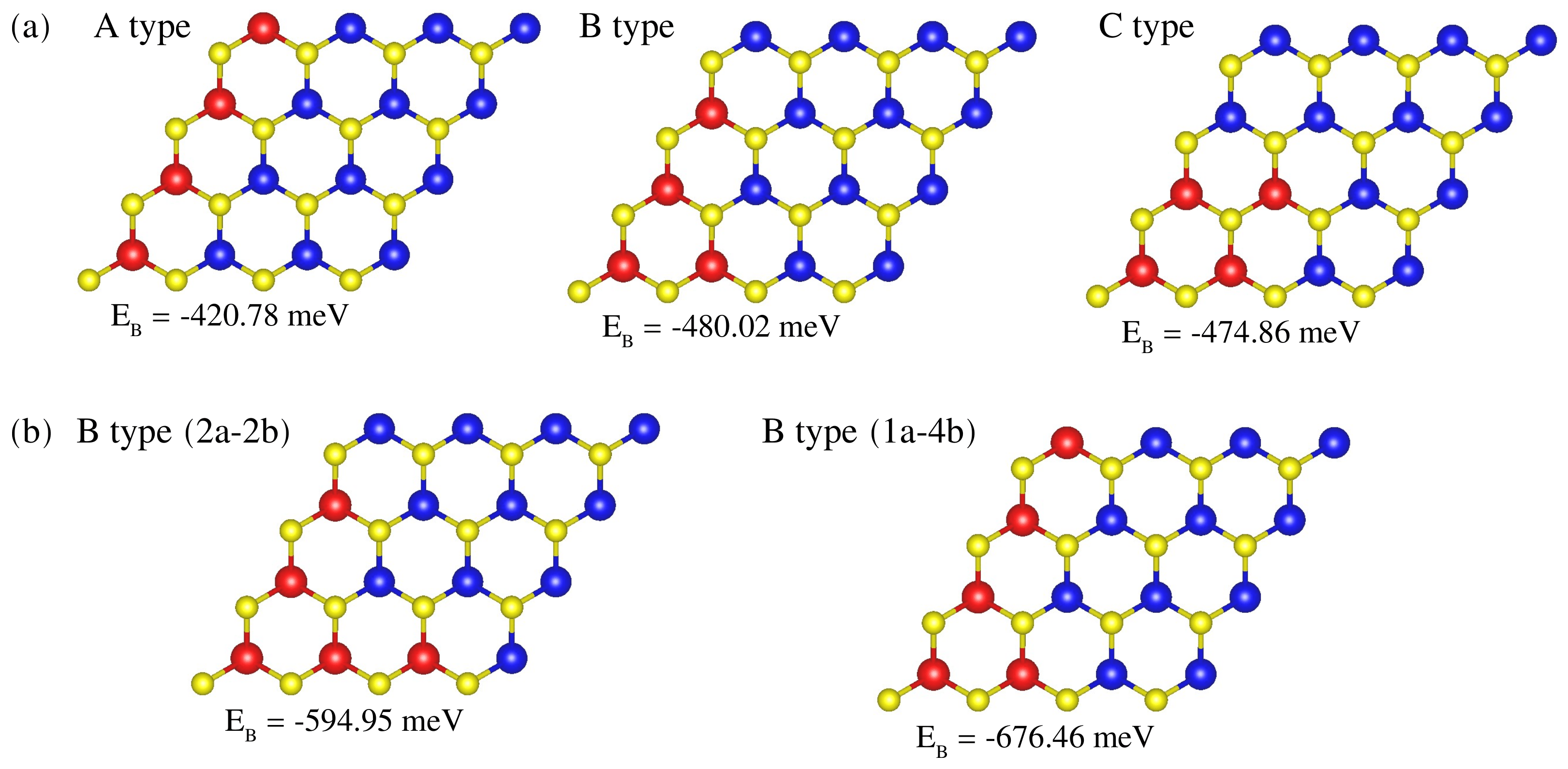}
\caption{ (a) Binding energy for different types of Nb dopant distribution in a 4 $\times$ 4 supercell. The shape of A, B and C type can be recognized as line, oblique-L and oblique-square, respectively. (b) Binding energy for two kinds of B type dopant distribution, where $\textbf{a}$ and $\textbf{b}$ refer to the two lattice vectors shown in Fig. 1(b). (Red: Nb, Blue: W, Yellow: S) }
\end{figure*}

\section{Results and Discussions}
\subsection{Energetic ordering of Nb cluster motifs}

From Figs. 1(a) and 1(b), in which only the W and Nb atoms can be seen, we see that the local atomic configurations of the Nb atoms (dark) in W$_{1-x}$Nb$_{x}$S$_{2}$ are quite varied, comprising motifs such as lines, clusters and isolated atoms. Judging by the small number of even darker sites, some vacancies may well be present, but it is unclear if these represent beam damage or are intrinsic, so we will not focus on them any further. We first study the relative stability of examplars of local arrangements of Nb atoms from the point of view of energetics, to inform us as to the situation that would be observed in thermal equilibrium.

Evaluating equation (2), we find the binding energy for two adjacent Nb atoms in WS$_{2}$ to be -125.79 meV.
This negative binding energy suggests that Nb atoms should tend to cluster within WS$_{2}$ if free to equilibrate. The situation is reversed if we consider the case of NbS$_{2}$ doped with W atoms, which produces a positive binding energy of 20.76 meV and suggests W atoms would not form clusters within a NbS$_{2}$ monolayer. For illustrative purposes, in Fig. S1 we show the result of a Monte Carlo simulation of W$_{0.9}$Nb$_{0.1}$S$_{2}$ assuming a pairwise interaction of -125.79 meV. For a large model cell (120$\times$120) in thermal equilibrium at 800 K, phase segregation is observed with two large Nb clusters forming. The contrast of this behaviour with that shown in Fig 1(a) demonstrates that kinetic effects during growth must control the observed dopant distribution.

Fig. 2 shows the binding energies for different designs of prototypical Nb dopant cluster in a 4 $\times$ 4 supercell. The energy differences are relatively small, but significant enough to conclude that less-linear forms such as B and C types are more energetically favorable than perfect lines such as A, with the oblique-L shape cluster emerging as the most stable form for 4 Nb atoms in this supercell. However, this trend does not continue to large scale: we classify more extended B-type clusters based on the length along the two lattice vectors (see Fig. 1(b)) in Fig. 2(b). Energetically an oblique-L with a longer line is the preferred structure for a 4 $\times$ 4 cell with 5 dopant atoms. Similarly for 6 atoms a line comprising two adjoining 3-membered triangles is favoured (see Fig. S2(a) in supplementary material). Isolated atoms (`point doping' following the nonmenclature of reference \onlinecite{Doping_in_TMDCs}) is the least favourable of all the tested configurations (Fig. S2(e) in supplementary material).

We verify the validity of our conclusions regarding energy ordering by modelling a selection of larger supercells: full results can be found in the supplementary materials and are summarised here. First, we extend along $\mathbf{b}$ to produce 12 $\times$ 4 supercells composed of combinations of the prototype Nb dopant distributions of Fig. 2. The binding energies of these models are shown in Fig. S3 and confirm the oblique-L as the favourable form for groups of 4 atoms. In Fig. S4, we show energetics for 6 $\times$ 6 supercells that show what happens on substitution of further Nb atoms into existing structures (4- and 5-membered lines). A perfect line spanning the periodic system is the lowest energy structure overall, and we see that the Nb atom favours completing an existing line rather than occupying any other site (compare Fig. S4(a) and (b) in supplementary material). However, if there is an isolated Nb atom as well as a 4-atom line, an additional Nb atom prefers to combine with the isolated atom than the line (compare Fig. S4(c) and (d) in supplementary material). This remains the case even if completing the line would make it periodic (though the difference is much reduced: compare Figs. S4(e) and S4(f)). This confirms the previous observation that two small lines or clusters are more energetically favourable than a single large line or cluster.

Taken together, these observations help explain why we see a self-limited length of the lines in the STEM (Fig. 1(a)), and why there are many small line segments with gaps between them, rather than an extended perfect line. Considerations of strain associated with the different sizes of Nb and W could contribute to explaining why these line segments would be aligned rather than random. However, overall, the observed occurrence of linear configurations cannot be explained purely through consideration of energetics: clearly these must combine with kinetic processes as the monolayer forms, and only jointly do these determine distributions, with the result that small clusters and lines are predominant.

\subsection{Effective bandstuctures for realistic model systems}

Using the spectral function unfolding method, we can calculate the effective bandstructure of large supercells with a range of the realistic atomic configurations.
Fig. 3 shows the effective bandstructures (EBS) along $\Gamma$ to M for four kinds of Nb dopant distribution, simulated in 12 $\times$ 12 supercells. The case named ``real'' is copied from a representative region of STEM image. A flat band above the Fermi level is seen in the ``perfect line'' and ``real'' cases. The flat band becomes less obvious in ``half line'' and then is duplicated in the ``random''. An obvious flat band may originate from a longer line (e.g. ``perfect line'' and ``real''). Note that there is no flat band for point doping \cite{Doping_in_TMDCs} indicating that flat bands are a phenomenon arising from localised states associated with extended clusters of dopants that strongly perturb the local electronic structure. A similar argument can be applied to the effective bandstructures along $\Gamma$ to K1, K2 and K3, as shown in Fig. 4. Here we observe that the topmost valence band crosses the Fermi level, indicating metallic behaviour for all Nb dopant distributions, associated with significant valence band splitting due to spin-orbit coupling. It first seems like our metallic results contrast to the results of Gao et al. \cite{Doping_in_TMDCs}. They showed that the material  still a semiconductor for 6.25\% Nb doping. This may be because a continuous transport channel cannot be formed at that concentration, since the length and separation of their line segments are shorter and larger than our ``half line'' case respectively, whereas the Nb composition in our ``random'' case (11.1\% Nb doping) is almost twice as large as in their calculation. Furthermore, the flat band in the bandstructure along $\Gamma$ to K1 in the ``perfect line'' case becomes a normal band (crossing the Fermi level).

We would naturally expect to observe a higher conductivity in the direction parallel to the line, resulting from the presence of extended states along that direction. The bandstructure of the ``real'' case lies somewhere between the two extreme cases of ``perfect line'' and ``random'' since the ``real'' distribution can be seen as being comprised of line, point and other distributions, with some gaps as in the ``half line'' case. The charge density distributions of the topmost bands in the ``real'' case are concentrated in the regions around Nb atoms (major) and W atoms (minor) with both mostly contributed by $d$ orbitals for VBM and the flat bands, while the CBM is mostly  contributed by the $d$ orbitals of W atoms, with a much smaller component on Nb atoms (Fig. S5 in supplementary material). It is worth noting that the charge distributions of the two supercell eigenstates most strongly associated with the two flat bands seen near $\Gamma$ in Fig. 4 are well-separated in space (hence their energy separation), whereas those at K are much more delocalised.

\begin{figure*}[bt]
  \centering
  \includegraphics[width=\textwidth]{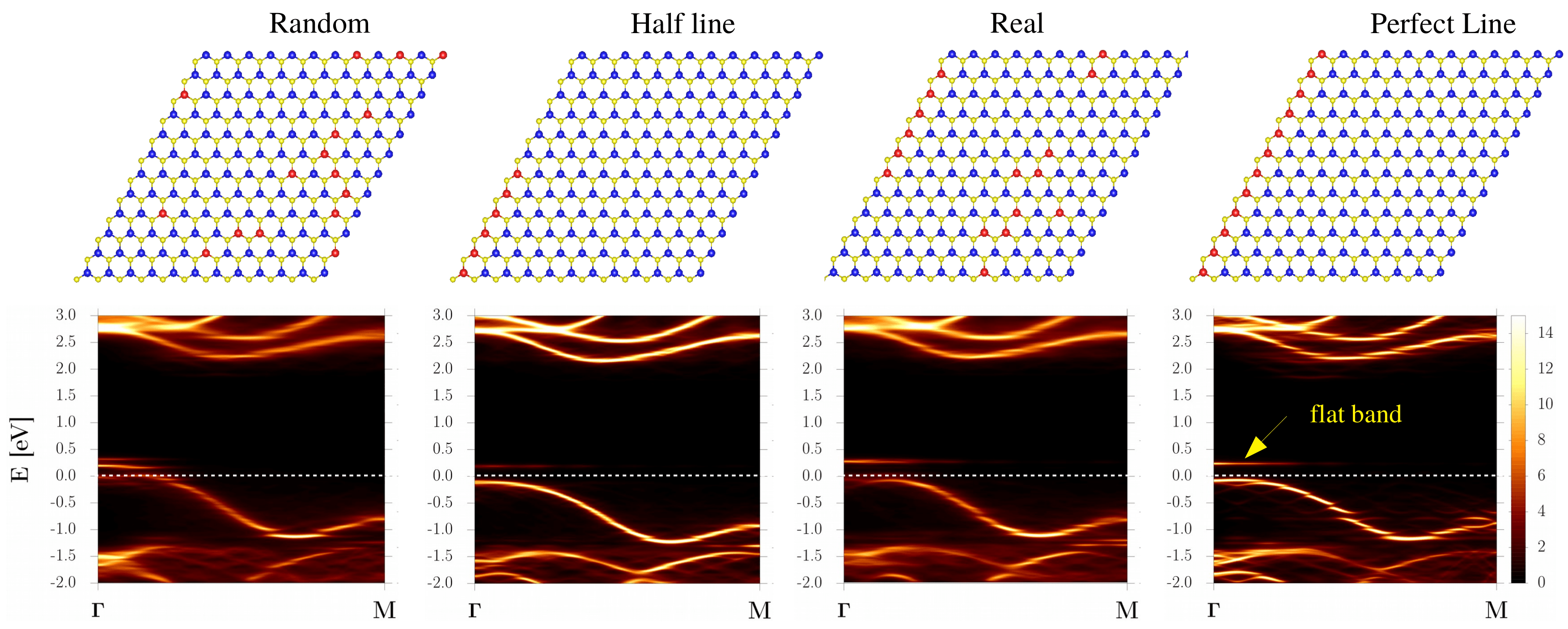}
\caption{Atomic configurations of four types of Nb dopant distribution (``random'', ``half line'', ``real'' and ``perfect line'') and their corresponding effective bandstructures along $\Gamma$ to M. Dashed line: Fermi level. (Red: Nb, Blue: W, Yellow: S) }
\end{figure*}
\begin{figure*}[hbt]
  \centering
  \includegraphics[width=\textwidth]{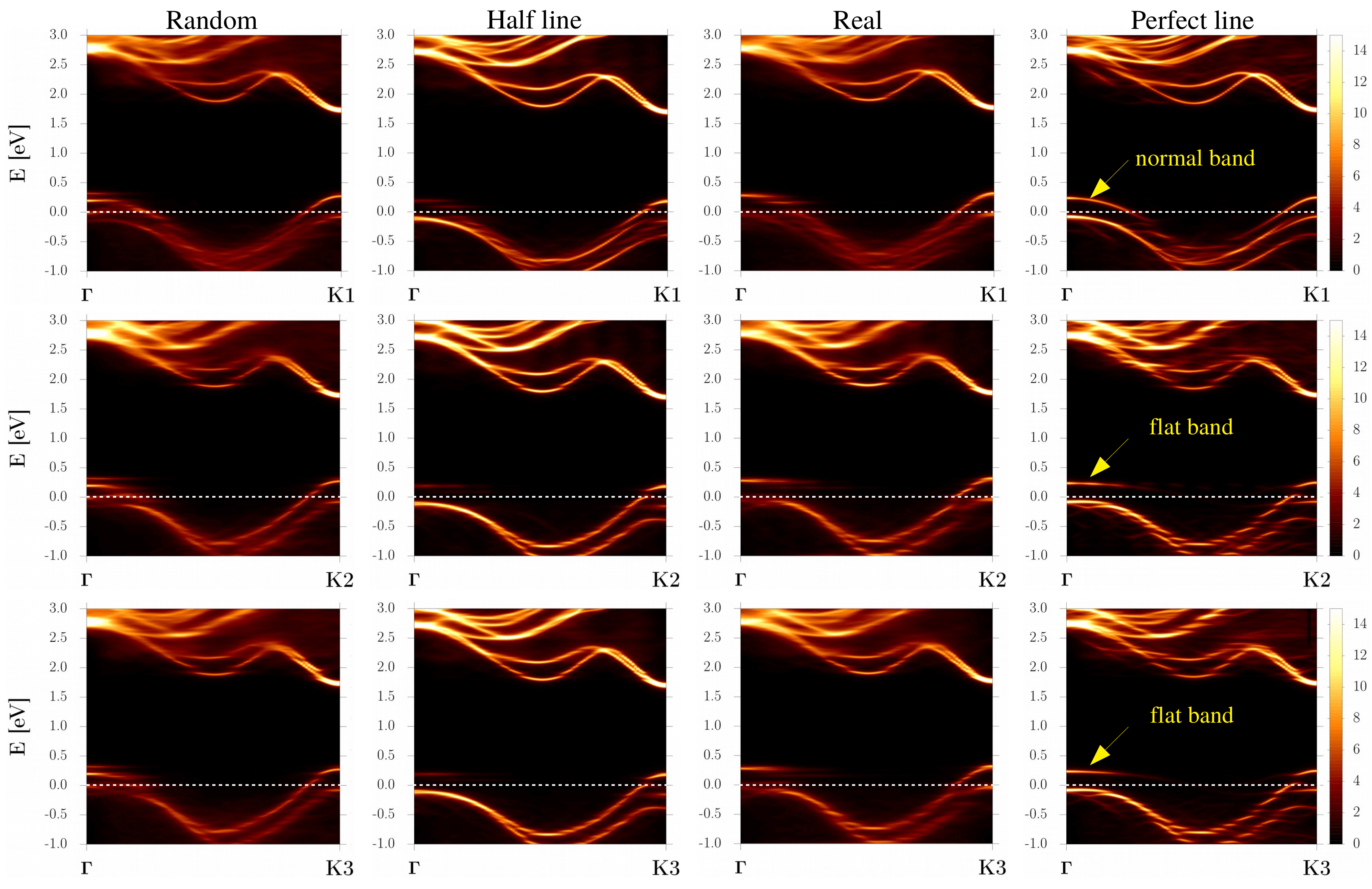}
\caption{Atomic configurations of four types of Nb dopant distribution (``random'', ``half line'', ``real'' and ``perfect line'') and their corresponding effective bandstructures along $\Gamma$ to (K1, K2 and K3) (defined in Fig. 1(c)). Dashed line: Fermi level. }
\end{figure*}

\subsection{Calculated semi-classical conductivities}

From Fig. 4, we observe a very interesting phenomenon in the effective bandstructure along $\Gamma$ to K1 for the ``perfect line'' case, namely that it appears close to exhibiting insulating behaviour along the $\Gamma$-K2 and $\Gamma$-K3 directions (no bands crossing the Fermi level, as the split-off bands are almost entirely separate) but is clearly metallic along $\Gamma$-K1 (i.e. for $k$-vectors aligned with the line there are bands crossing the Fermi level). Therefore, we now calculate and compare the conductivity of the four types of Nb dopant distributions along two real-space directions: parallel to the dopant line and perpendicular to the dopant line, as indicated schematically by means of green arrows in Fig. 5. In Fig. 5, the ``real (line Nb)'' and ``real (isolated Nb)'' are based on two hand-chosen, but nevertheless representative, $8\times 8$-cell regions from the STEM image in Fig. 1(a). The calculated conductivities, expressed in units of the relaxation time $\tau$, are around 10$^{18}$. The conductivity is plotted against $n$, representing a chosen level of extrinsic carrier doping which moves the Fermi level. For an alloy sample with a carefully chosen Nb content and no other intentional doping, the most relevant value is around $n=0$ but for context we show a range of doping concentration from $-5\times 10^{14}$ cm$^{-2}$  to $5\times 10^{14}$  cm$^{-2}$.

We find, as anticipated, that it is possible to observe highly anisotropic conductivity, especially for the ``perfect line'' case, but also to a significant extent for the ``real (isolated Nb)'', ``real (line Nb)'' and ``half line'' cases. For all four cases, the conductivity along the direction parallel to the line is substantially higher than the conductivity along the direction perpendicular to the line, resulting in a ratio ranging from 1.6 (``real (isolated Nb)'') to 17.2 (``perfect line'') at the peak, which in all those cases occurs very close to zero extrinsic doping. Here, extrinsic doping refers to dopant charge arising from any species other than the Nb arising from the chosen alloy concentration $x$.  Although the concentration of Nb atoms in the ``random'' ($x$= 0.219) and ``real (line Nb)'' ($x$= 0.141) models are higher, while for ``real (isolated Nb)'' it is the same as the ``perfect line'' ($x$= 0.125), the highest anisotropy ratio happens at the ``perfect line'' case at zero extrinsic doping with a value of about 17.2. This may be because holes can move much more easily in the direction parallel to the line compared with other dopant distributions. Combined, these observations strongly indicate that the real samples would display a degree of anisotropy somewhere in the range exhibited by the samples measured here.

Notably, the second highest ratio of the modelled systems, at exactly zero extrinsic doping, is the ``real (line Nb)'' case (ratio $\simeq$ 5.0). One can infer that the conductivity ratio in the ``real (line Nb)'' case is higher than in the ``half line'' ($\simeq$ 1.8) case because the former includes more continuous Nb line than the latter: specifically, two gaps of one site reduces the matrix element for hopping by less than one gap of 4 sites does, for these models. The ``perfect line'' (3.72 $\times$ 10$^{18}$ \: $\Omega^{-1}$ m$^{-1}$ s$^{-1}$) has only twice as large a conductivity as the ``real (line Nb)'' case (1.89 $\times$ 10$^{18}$ \: $\Omega^{-1}$ m$^{-1}$ s$^{-1}$) along the direction parallel to the line at zero doping concentration (although the ``perfect line'' model has a slightly lower Nb concentration than the ``real (line)'') . This implies that the influence on conductivity of the extra Nb doping is much stronger than the limitations imposed by hopping of holes. Meanwhile, the conductivity of the ``perfect line'' model, along the direction perpendicular to the line, is 57\% that of the ``real (line Nb)'', resulting in an anisotropy ratio 3.4$\times$ larger for the ``perfect line'' case compared to the ``real (line Nb)'' case. The anisotropy ratio of the ``random'' model is of course the lowest: by chance the value is below 1 (ratio $\simeq$ 0.7) for this particular realisation. Therefore, we conclude that the conductivity and its anisotropy can be very significantly increased by increasing the strength and degree of geometric perfection of the line formation in the WS$_{2}$ monolayer, and is also influenced by the total concentration of Nb dopant atoms. It remains to be seen how perfect a line can be formed in real systems, given the limitations of the growth processes, but the fact that the ``real'' models show high anisotropy is very significant.

\begin{figure*}[bt]
  \centering
  \includegraphics[width=\textwidth]{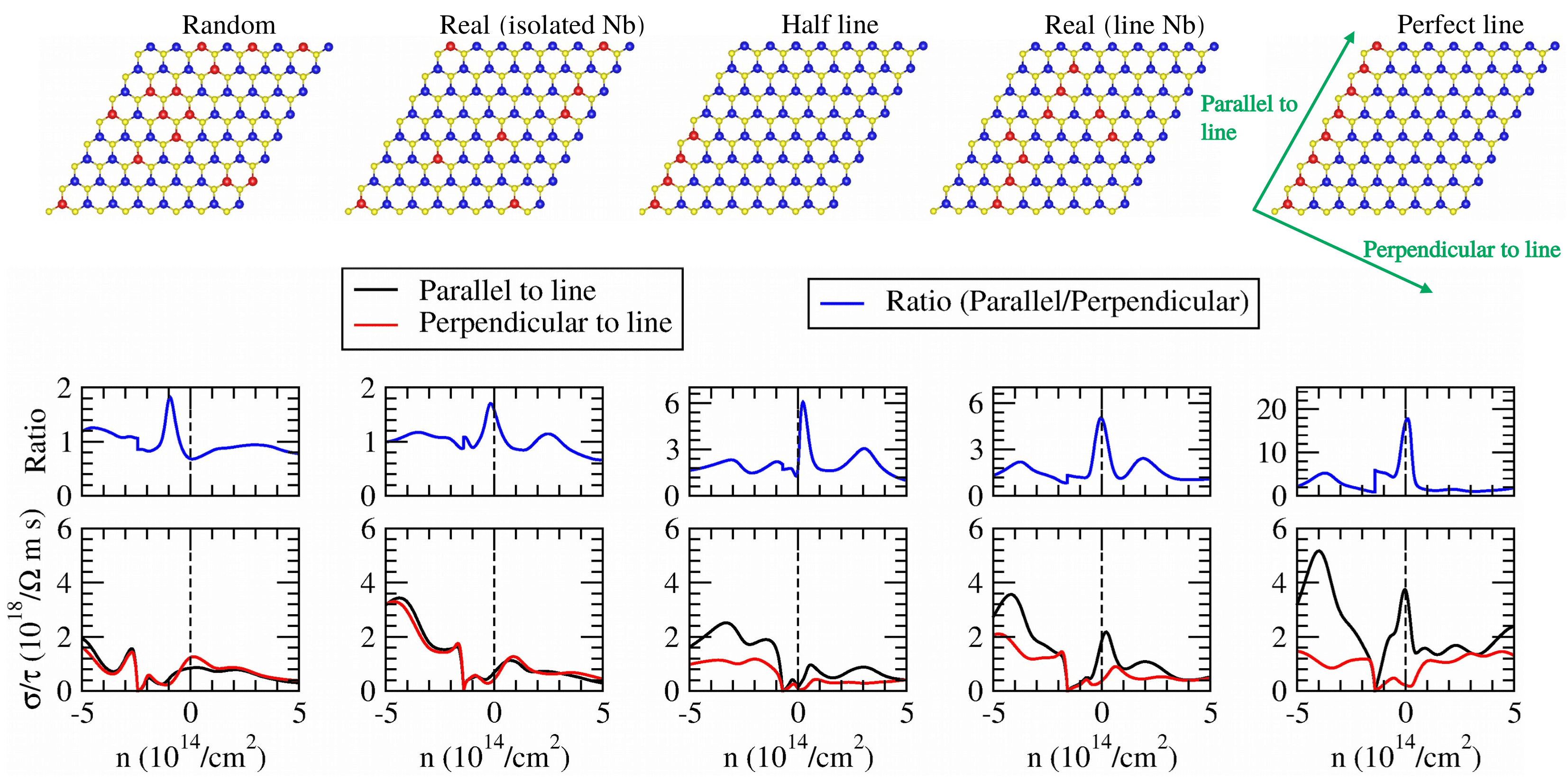}
\caption{Top panels show the atomic configurations of five types of Nb dopant distribution: ``random'' ($x$ $=$ 0.219), ``real (isolated Nb)'' ($x$ $=$ 0.125), ``half line'' ($x$ $=$ 0.063), ``real (line Nb)'' ($x$ $=$ 0.141) and ``perfect line'' ($x$ $=$ 0.125) (Red: Nb, Blue: W, Yellow: S). Bottom panels show their corresponding conductivities (in units of relaxation time) along the direction of parallel to the line and perpendicular to the line as a function of doping concentration at T $=$ 300 K. The middle panels (blue lines) show the ratio of the conductivity along the direction parallel to the line to that in the direction perpendicular to the line, to clearly illustrate the anisotropy.  }
\end{figure*}

\section{Conclusions}

We observe highly anisotropic distributions in samples of an examplar TMDC subjected to aliovalent doping, namely W$_{1-x}$Nb$_{x}$S$_{2}$. We examine using theoretical methods the consequences this has for bandstructure and conductivity. The anisotropic configuration, in particular the highly-coherent line structures observed in our as-grown samples, have been partly explained from the point of view of energetics, yet kinetic effects must still play an important role in the formation mechanism of the specific patterns of dopant distribution, particularly in regards to symmetry-breaking by the growth front. The effective bandstructure and the conductivity both reflect the anisotropy of the dopant distribution. Furthermore, a very high level of anisotropy in the conductivity can be realised by increasing the length, spacing and degree of perfection of the line pattern of substitional dopants, as well as by the overall dopant composition. While the exact degree of anisotropy will depend on the dopant distribution, in our model systems we find conductivity ratios as high as 5:1 for realistic models of W$_{1-x}$Nb$_{x}$S$_{2}$ with $x$ in the range 0.05 to 0.22. Crucially, it is not necessary to invoke extended ``perfect'' lines of dopants, which are in any case not observed experimentally, in order to predict high levels of in-plane anisotropy in the conductivity. 

Our study is restricted to the W$_{1-x}$Nb$_{x}$S$_{2}$ monolayer in this work, but the similarities of the bandstructure among this class of materials means it is reasonable to expect similar behaviour for any isoelectronic system of the form M$_{1-x}$D$_{x}$X$_{2}$ where M$=$Mo, W and X$=$S, Se, and D is an appropriately-chosen aliovalent dopant atom, such as Nb or Ta, introducing shallow holes. It is also reasonable to expect related behaviour in electron-doped systems, though the details may well vary due to the different arrangement of the bands at the conduction band minimum for TMDCs. Energetic considerations indicate that it may not be feasible to grow extended ``perfect'' lines of substitutional dopants in such materials, but that the electronic transport properties even of samples incorporating realistic impurity distributions observed in STEM, may be very highly anistropic in spite of their imperfections.

\section{Acknowledments}

The Engineering and Physical Sciences Research Council supported NDMH and NRW through grant EP/P01139X/1.
SML and XX were supported by University of Warwick Chancellor’s Scholarships. Computing resources were provided by the Scientific Computing Research
Technology Platform of the University of Warwick, and the UK national high performance computing service, ARCHER, via the UKCP consortium (EP/P022561/1). We acknowledge the use of Athena at HPC Midlands+, which was funded by the EPSRC through Grant No. EP/P020232/1.

\end{document}


\title{Supporting Information for: Strong In-plane Anisotropy in the Electronic Properties of Doped Transition Metal Dichalcogenides exhibited in W$_{1-x}$Nb$_{x}$S$_{2}$}

\author{Siow Mean Loh}
\thanks{These authors contributed equally.}
\affiliation{Department of Physics, University of Warwick, Coventry CV4 7AL, United Kingdom.}

\author{Xue Xia}
\thanks{These authors contributed equally.}
\affiliation{Department of Physics, University of Warwick, Coventry CV4 7AL, United Kingdom.} 

\author{Neil R. Wilson}
\affiliation{Department of Physics, University of Warwick, Coventry CV4 7AL, United Kingdom.} 

\author{Nicholas D. M. Hine}
\email{n.d.m.hine@warwick.ac.uk}
\affiliation{Department of Physics, University of Warwick, Coventry CV4 7AL, United Kingdom.} 

\date{\today}


\maketitle

\section{Supporting information}
\renewcommand{\thefigure}{S\arabic{figure}}
\renewcommand{\thetable}{S\Roman{figure}}

Fig. \ref{fig:S1} and Table \ref{table:1} shows the energetic ordering of the dopant distributions in WS$_{2}$ monolayer within a 4$\times$4 supercell. The ordering is:
(a) Two small triangular clusters $>$ A big triangular cluster $>$ A big oblique-rectangular cluster \\
(b) Line $>$ Isolated \\
(c) Closed-line $>$ Separated-line (not seemed in the real atomic configutation due to the absence of the kinetic effect)

In Fig. \ref{fig:S2} and Table \ref{table:2} we show that the energetic ordering of the dopant distributions as calculated in the small supercell is preserved upon moving to a larger supercell comprising the combination of two types of Nb dopant distribution.

In Fig. \ref{fig:S3} and Table \ref{table:3} we demonstrate, via calculations in a 6$\times$6 supercell, that the following are true:\\
(a) If 5 Nb atoms form a line, with no other Nb present, a further Nb atom prefers to combine with the line rather than occupy any other site (compare (a) vs (b)). \\
(b) If 5 Nb atoms form a line and there is an isolated Nb atom elsewhere, the next Nb atom prefers to combine with the isolated Nb atom rather than combining with the line (compare (c) vs (d)). \\
(c) If 5 Nb atoms form a line and there is an isolated Nb atom elsewhere, the next Nb atom prefers to combine with the isolated Nb atom to make a new line segment, rather than completing line such that it spans the whole supercell  (compare (e) vs (f)). \\
Overall, from the perspective solely of energetics, linear motifs tend to be the most favourable for a given number of Nb dopants. However, the real behaviour must still depend on kinetic factors, particularly as a source of symmetry-breaking.

Fig. \ref{fig:S4} shows the charge density distributions of chosen bands of the supercell for the real atomic configurations.

\newpage

\setcounter{figure}{0}    
\begin{figure*}
  \centering
\includegraphics[width=\textwidth]{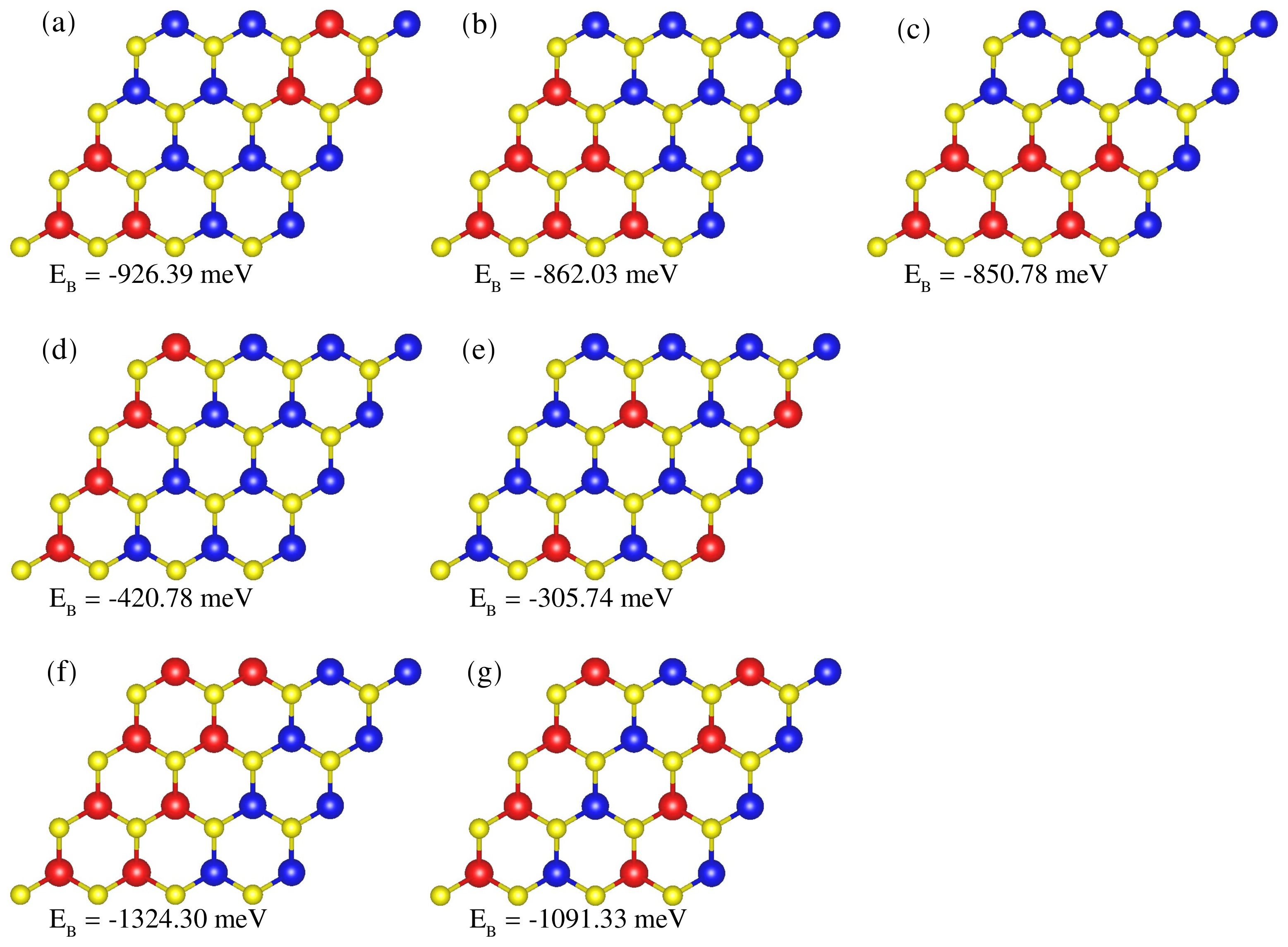}
    \caption{The binding energy, calculated in a 4 $\times$ 4 supercell, of (a) two small triangular clusters (b) a big triangular cluster (c) a big oblique-rectangle cluster (d) line (e) isolated (f) closed-line (g) separated-line Nb dopant distributions. The lowest binding energy are the two small triangular clusters, line and closed-line among [(a), (b) and (c)], [(d) and (e)] and [(f) and (g)], respectively. (Red: Nb, Blue: W, Yellow: S) }\label{fig:S1}
\end{figure*}

\begin{table}[h!]
\begin{tabular}{ |c|c| }
\hline
(a) & -926.39 meV \\
(b) & -862.03 meV \\
(c) & -850.78 meV \\
\hline
(d) & -420.78 meV \\
(e) & -305.74 meV \\
\hline
(f) & -1324.30 meV \\
(g) & -1091.33 meV \\
\hline
\end{tabular}
\caption{ The binding energies of each of the dopant distributions in Fig. S1. }
\label{table:1}
\end{table}

\setcounter{figure}{1}    
\begin{figure*}
  \centering
\includegraphics[width=\textwidth]{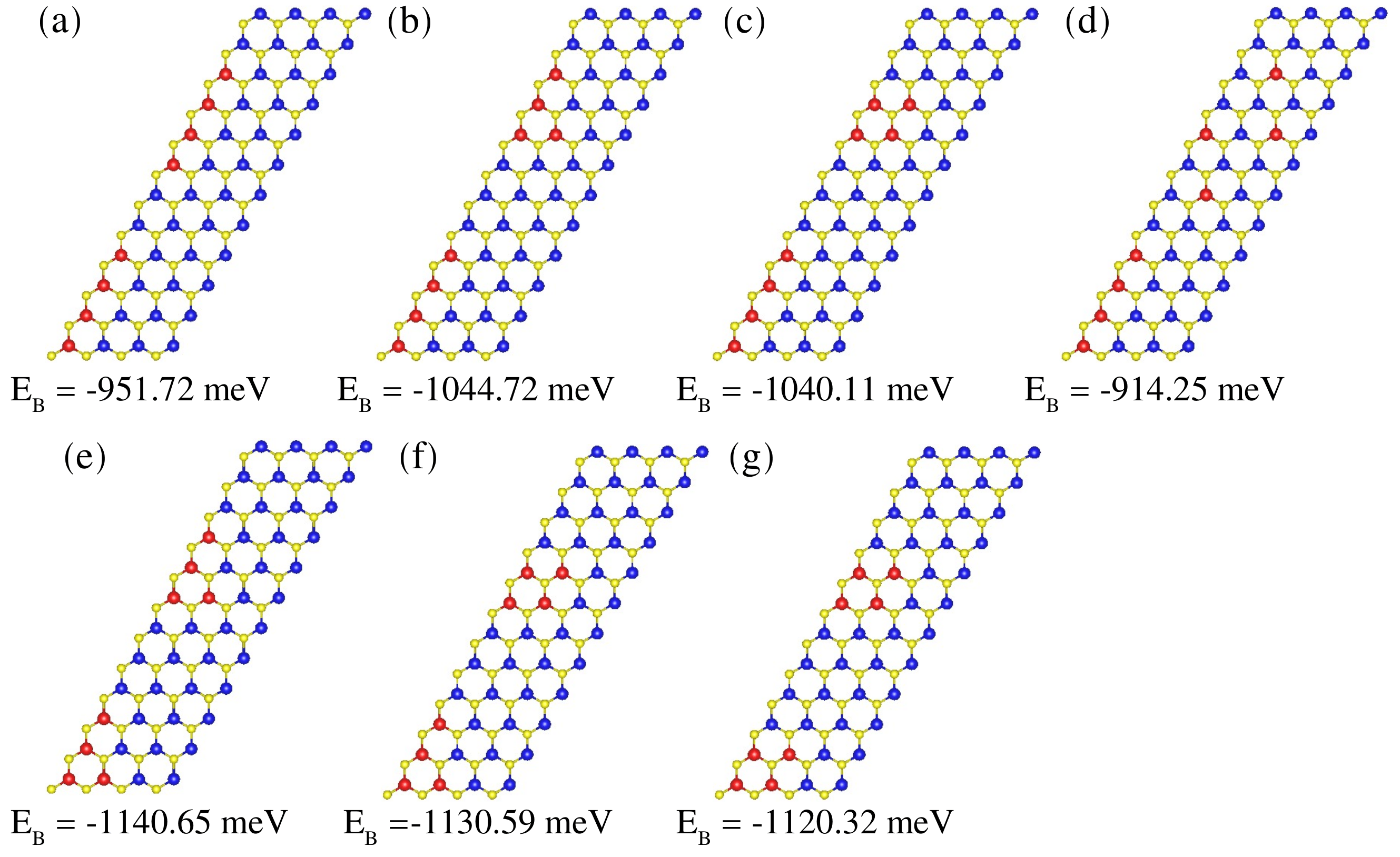}
    \caption{The combination of two types of Nb dopant distributions in a larger supercell (4 $\times$ 12 supercell). (a) Line + Line (b) Line + Oblique-L (c) Line + Oblique-Square (d) Line + Isolated (e) Oblique-L + Oblique-L (f) Oblique-L + Oblique-Square (g) Oblique-Square + Oblique-Square. The lowest binding energy is the ``Oblique-L + Oblique-L'' case. (Red: Nb, Blue: W, Yellow: S) }\label{fig:S2}
\end{figure*}

\begin{table}[h!]
\begin{tabular}{ |c|c| }
\hline
(a) & -951.72 meV \\
(b) & -1044.72 meV \\
(c) & -1040.11 meV \\
(d) & -914.25 meV \\
(e) & -1140.65 meV \\
(f) & -1130.59 meV \\
(g) & -1120.32 meV \\
\hline
\end{tabular}
\caption{ The binding energy of each of the dopant distributions in Fig. S2. }
\label{table:2}
\end{table}

\setcounter{figure}{2}

\begin{figure*}
  \centering
\includegraphics[width=0.93\textwidth]{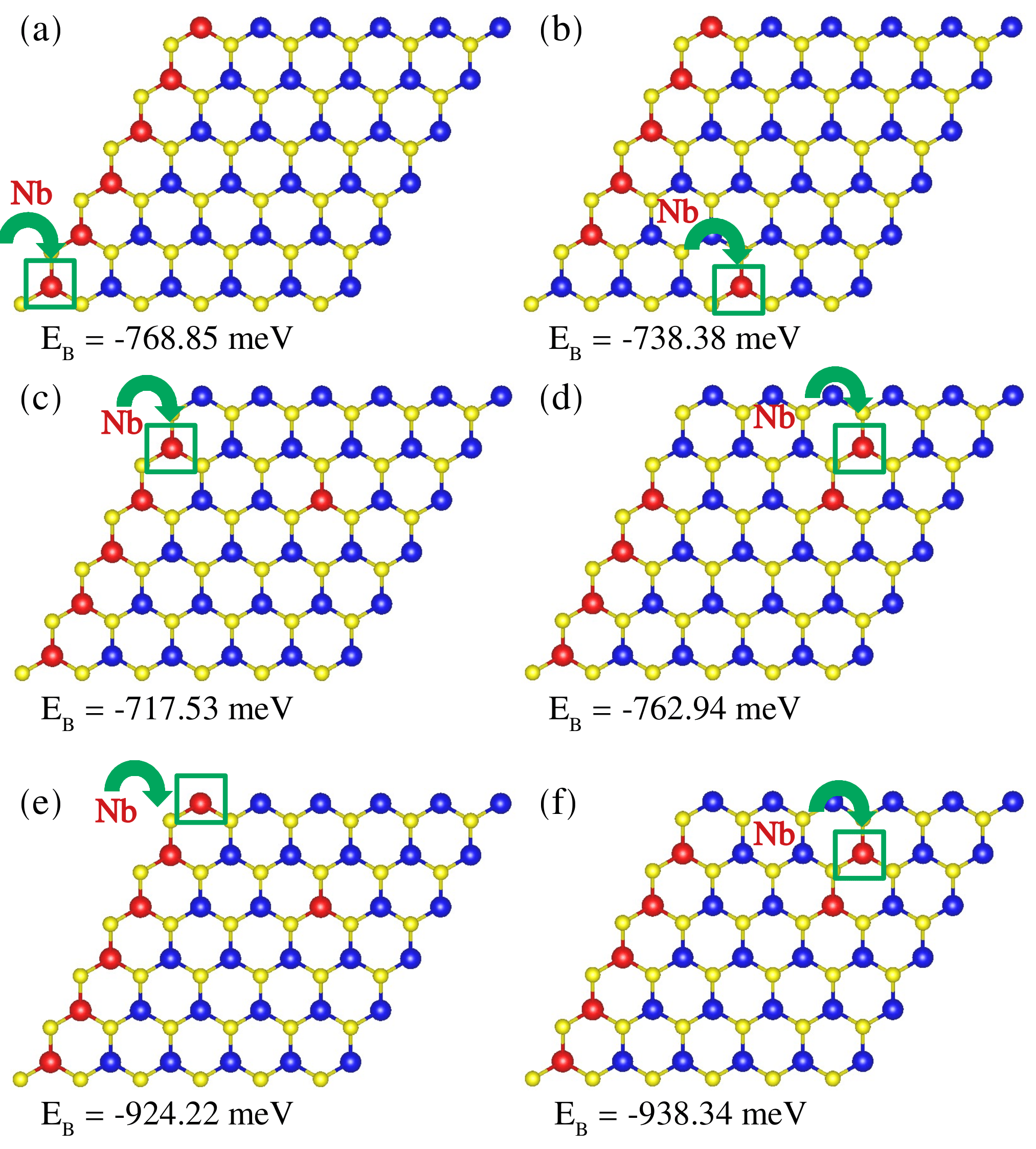}
    \caption{Energetics of various Nb dopant distributions in the 6 $\times$ 6 supercell. \\ 
(a) Line (6 Nb atoms) vs. (b) Line (5 Nb atoms) + Point  ((a) has lower binding energy) \\ 
(c) Line (5 Nb atoms) + Point vs. (d) Line (4 Nb atoms) + Line (2 Nb  atoms)  ((d) has lower binding energy) \\ 
(e) Line (6 Nb atoms) + Point vs. (f) Line (5 Nb atoms) + Line (2 Nb atoms).  ((f) has lower binding energy) \\
(Red: Nb, Blue: W, Yellow: S) }\label{fig:S3}
\end{figure*}

\begin{table}[ht]
\begin{tabular}{ |c|c| }
\hline
(a) & -768.85 meV \\
(b) & -738.38 meV \\
\hline
(c) & -717.53 meV \\
(d) & -762.94 meV \\
\hline
(e) & -924.22 meV \\
(f) & -938.34 meV \\
\hline
\end{tabular}
\caption{ The binding energy of each of the dopant distributions in Fig. S3. }
\label{table:3}
\end{table}

\setcounter{figure}{3}    
\begin{figure*}
  \centering
\includegraphics[width=\textwidth]{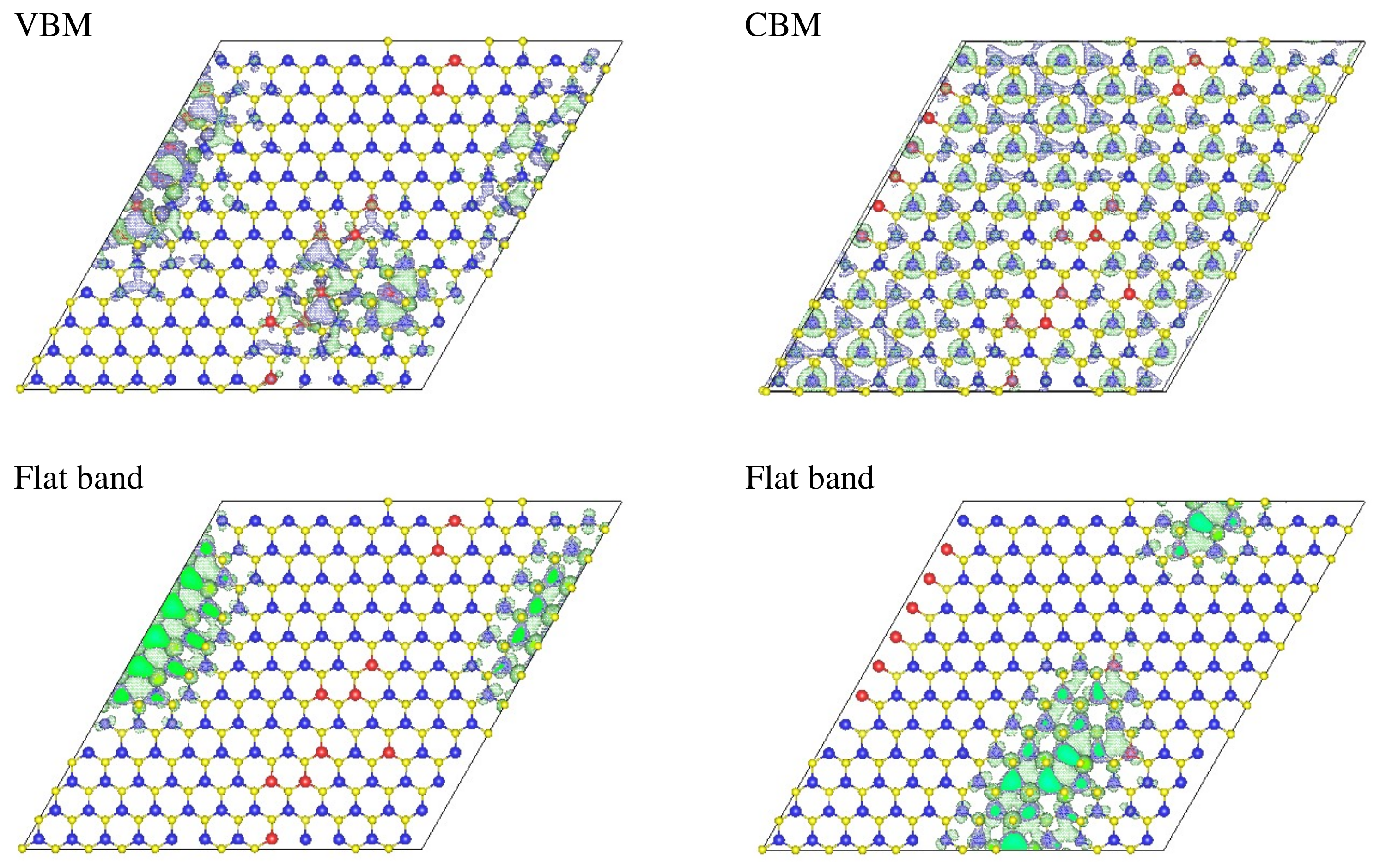}
    \caption{ Charge density distribution of real atomic configurations for the bands below the Fermi level (VBM) and above the Fermi level (CBM) and around the flat bands in Fig. 3 and Fig. 4. The charge density is mostly contributed by the $d$ orbitals of Nb atoms (major) and W atoms (minor) for VBM and the flat bands. Meanwhile for the CBM, the majority and minority contributions come from the $d$ orbitals of W atoms and Nb atoms respectively.  It is worth noting that the charge distributions of the two supercell eigenstates most strongly associated with the two flat bands seen near $\Gamma$ in Fig. 4 are well-separated in space (hence their energy separation), whereas those at K are much more delocalised.  (Red: Nb, Blue: W, Yellow: S)}\label{fig:S4}
\end{figure*}